\documentclass[reprint,
superscriptaddress,
frontmatterverbose,
showpacs,preprintnumbers,
 amsmath,amssymb,
 aps,pra,Ifloatfix,
citeautoscript,
]{revtex4-2}

\usepackage{mdframed}

\usepackage[pdftex]{graphicx}

\usepackage{mathrsfs}
\usepackage[linkcolor=blue,colorlinks=true,citecolor=blue,filecolor=blue]{hyperref}
\usepackage{amsmath}
\usepackage[english]{babel}
\usepackage{booktabs}
\usepackage{amssymb}
\usepackage{type1cm}
\usepackage{braket}
\usepackage{tikz}
\usepackage{pgfplots}
\usepackage{lipsum}
\usepackage{xcolor}
\usepackage{graphicx}
\usepackage{subcaption}
\usepackage{dcolumn}
\usepackage{bm}
\usepackage{physics}
\usepackage[utf8]{inputenc}
\usepackage{multirow}
\usepackage{tikz}

\definecolor{Midnight_Blue}{rgb}{0.1, 0.1, 0.6}

\usepackage{cleveref}
\crefname{equation}{Eq.}{Eqs.}
\Crefname{equation}{Equation}{Equations}
\crefname{figure}{Fig.}{Figs.}
\Crefname{figure}{Figure}{Figures}
\crefname{figure}{Fig.}{Figs.}
\Crefname{figure}{Figure}{Figures}
\crefname{section}{Supplemental Material Section}{Supplemental Material Sections}
\Crefname{section}{Supplemental Material Section}{Supplemental Material Sections}
\crefname{appendix}{Appendix}{Appendices}
\Crefname{appendix}{Appendix}{Appendices}
\crefname{table}{Table}{Tables}
\Crefname{table}{Table}{Tables}

\usepackage{enumitem,amssymb}
\newlist{todolist}{itemize}{2}
\setlist[todolist]{label=$\square$}
\usepackage{pifont}
\usepackage{etoolbox}
\pgfplotsset{compat=1.18} 
\begin{document}

\title{Enhancing Long-distance Continuous-variable Quantum-key-distribution with an Error-correcting Relay}

\author{S. Nibedita Swain}
\email{swain.snibedita@gmail.com}
\affiliation{School of Mathematical and Physical Sciences,
University of Technology Sydney, Ultimo, NSW 2007, Australia}

\affiliation{Sydney Quantum Academy, Sydney, NSW 2000, Australia}
\affiliation{Centre for Quantum Computation and Communication Technology, School of Mathematics
and Physics, University of Queensland, Brisbane, Queensland 4072, Australia}
\author{Ryan J. Marshman}
\affiliation{Centre for Quantum Computation and Communication Technology, School of Mathematics
and Physics, University of Queensland, Brisbane, Queensland 4072, Australia}

\author{Josephine Dias}
\affiliation{Centre for Quantum Computation and Communication Technology, School of Mathematics
and Physics, University of Queensland, Brisbane, Queensland 4072, Australia}

\author{Matthew S. Winnel}
\affiliation{Centre for Quantum Computation and Communication Technology, School of Mathematics
and Physics, University of Queensland, Brisbane, Queensland 4072, Australia}

\author{Alexander S. Solntsev}
\affiliation{School of Mathematical and Physical Sciences,
University of Technology Sydney, Ultimo, NSW 2007, Australia}

\author{Timothy C. Ralph}
\affiliation{Centre for Quantum Computation and Communication Technology, School of Mathematics
and Physics, University of Queensland, Brisbane, Queensland 4072, Australia}

\begin{abstract}
   Noiseless linear amplifiers (NLAs) serve as an effective means to enable long-distance continuous-variable (CV) quantum key distribution (QKD), even under realistic conditions with non-unit reconciliation efficiency. Separately, unitary averaging has been suggested to mitigate some stochastic noise, including phase noise in continuous-variable states. In this work, we combine these two protocols to simultaneously compensate for thermal-loss effects and suppress phase noise, thereby enabling long-distance CV QKD that surpasses the repeaterless bound, the fundamental rate-distance limit, for repeaterless quantum communication systems.

\end{abstract}

\date{\today}

\frenchspacing

\maketitle

\section{Introduction} 

Quantum Key Distribution (QKD) enables the creation of a shared, secure key between a sender and a receiver, even when communication occurs over an untrusted channel that may be entirely controlled by an eavesdropper. Within this context, continuous-variable (CV) QKD has emerged as a particularly promising approach, both from theoretical and experimental perspectives \cite{weedbrook2012gaussian}. In CV QKD schemes \cite{ralph1999continuous, grosshans2002continuous, weedbrook2004quantum, grosshans2005collective}, the sender (Alice) encodes information onto the quadratures of a optical field, typically using Gaussian modulations, which is subsequently transmitted through a channel to the receiver (Bob). Bob then performs either homodyne or heterodyne (double-homodyne) measurements \cite{laudenbach2018continuous}. The secure key is derived following a reconciliation procedure in which one of the communicating parties raw key is treated as the correct key: direct reconciliation \cite{grosshans2003virtual} for Alice's key, and reverse reconciliation if it is Bob's key. Using reverse reconciliation allows a non-zero secure key rate for any transmission distance in the asymptotic limit of a pure-loss channel \cite{grosshans2002reverse,lodewyck2007quantum} and so is what is typically considered.

In practical scenarios, the secret key rate (SKR), defined as the number of bits of secret key shared between Alice and Bob per use of the quantum channel, is significantly influenced not only by loss and thermal noise but also by phase noise. These factors both severely impact the achievable SKR and limit the transmission distance at which the SKR ultimately vanishes. Whilst thermal-loss channels have long been analysed, a recent study by Kish et al., demonstrated that phase noise notably degrades the performance of CV-QKD systems \cite{kish2022comparison}. Here, we propose to address these challenges by modifying the standard no-switching, Gaussian modulation protocol by introducing strategies that effectively mitigate both phase noise and thermal loss in order to extend the maximum possible transmission distance. Two promising techniques in this regard are unitary averaging (UA) \cite{marshman2018passive, swain2024improving} and heralded noiseless linear amplification (NLA) \cite{ralph2009nondeterministic, ralph2011quantum, xiang2010heralded, adnane2019quantum}, both of which offer intriguing avenues to enhance the robustness and range of CV-QKD implementations.

UA is a framework that is effective in mitigating phase errors in CV linear optical systems \cite{swain2024improving} albeit at the cost of making the channel probabilistic. In such setups, each input state is simultaneously sent through multiple independent copies of the transmission channel before being recombined with vacuum heralding of the non-output channels. Previous results \cite{swain2024improving} have indicated this process provides substantial benefits in protecting Gaussian states against phase errors. Notably, the robustness of this protocol persists even in the presence of losses within the optical components.

An ideal. probabilistic NLA with amplitude gain $g$ can extend the maximum transmission distance in proportion to $\log g $ \cite{blandino2012improving}. However, any physically realizable NLA can only approximate the ideal performance for optical signals with low amplitudes. Recently, CV QKD schemes employing Quantum Scissors (QS) \cite{ralph2008nondeterministic} to act as a Noiseless Linear Amplifier (NLA) in the low photon number regime have been investigated, enabling long-distance CV-QKD under conditions of sufficiently low channel excess noise \cite{ghalaii2020discrete}. Winnel et al., demonstrated that QS devices can function as loss-tolerant quantum relays by themselves, simultaneously performing entanglement distillation and entanglement swapping, effectively acting as a distributed NLA. Consequently, the repeaterless, or Pirandola–Laurenza–Ottaviani–Banchi (PLOB), bound \cite{pirandola2017fundamental} can be surpassed without the need for quantum memories \cite{winnel2021overcoming}.

In this work, we first demonstrate that, despite its probabilistic nature, UA can increase the transmission distance of CV-QKD in the presence of phase-noise. We then study how it can be successfully combined with a distributed NLA  relay scheme.
The first half of the article focuses on modelling phase noise within the entanglement-based (EB) version of the CV-QKD protocol and implement the UA protocol to demonstrate its improvement to secret key rate.
In the second half, we consider the effect of thermal loss \emph{and} phase noise on the performance of a distributed NLA based repeater set-up. We then combine UA an NLA, forming a hybrid protocol capable of surpassing the PLOB bound and enabling long-distance CV-QKD even in the presence of substantial noises.

The structure of this paper is as follows. Section \ref{THE ORIGINAL QKD PROTOCOL} reviews no switching CV-QKD protocol we will be using throughout. Section \ref{UA-Assisted CV QKD} presents how phase noise constrains the transmission distance and then introduces the UA-assisted CV-QKD protocol. In Section \ref{Hybrid UA-QS CV QKD}, we first discuss the impact of phase noise in the distributed NLA scheme, after which we present the hybrid UA–NLA protocol, including a comparison of the SKRs of the protocols considered. Finally, Section \ref{Conclusion} summarizes our results and offers some concluding remarks.

\section{THE CV QKD PROTOCOL}
\label{THE ORIGINAL QKD PROTOCOL}
\begin{figure*}

    \begin{subfigure}{2\columnwidth}
        \includegraphics[scale=0.6]{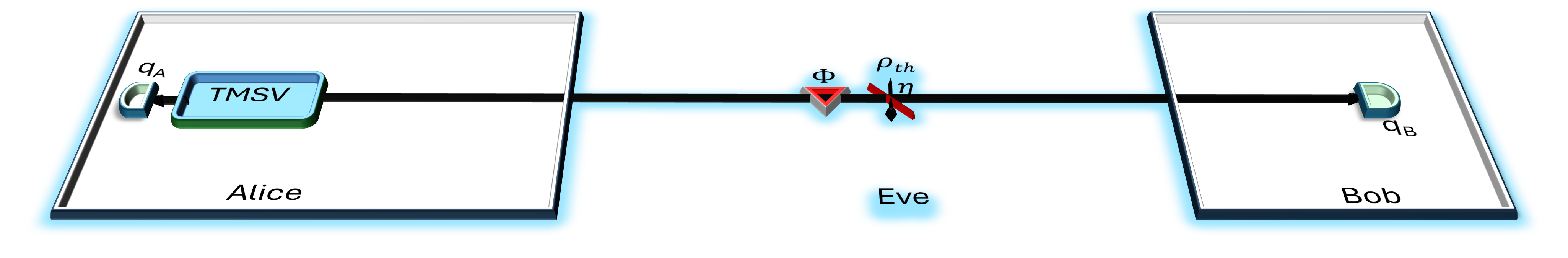}
        \caption{Entanglement based no-switching protocol.\label{fig:CirPh}}
    \end{subfigure}
    \begin{subfigure}{2\columnwidth}
		\includegraphics[scale=0.6]{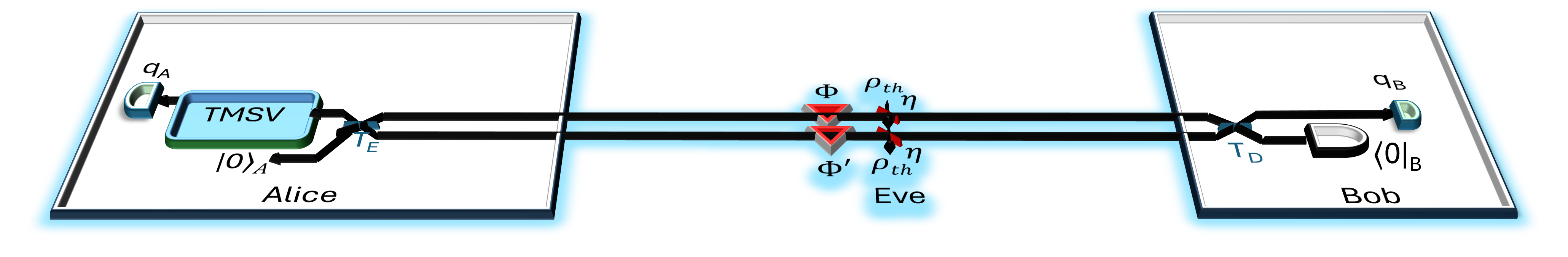}
    \caption{Entanglement based no-switching protocol with UA.\label{fig:CIRUA}}
  \end{subfigure}
\caption{The entangelment based no-switching protocol with a phase noise ($\phi$) and thermal ($\rho_{th}$) loss ($\eta$) channel both without \ref{fig:CirPh} and with \ref{fig:CirPh} unitary averaging. TMSV denotes the two-mode squeezed vacuum state. $q_A$ and $q_B$ mark the heterodyne detectors of Alice and Bob respectively. Alice prepares a TMSV state and, when utilising UA, she then combines the second mode of the TMSV state with the vacuum mode using a (50{:}50) encoding beamsplitter ($T_{E}$). The signal is then transmitted through the phase and thermal-loss channel(s) and, if necessary is subsequently recombined by Bob using a (50{:}50) decoding beamsplitter ($T_{D}$.). Bob then performs a heterodyne measurement on this received mode to obtain the resulting noisy key.} \label{fig:Cir Ph and UA}
\end{figure*}

\begin{figure*}[!htb]
	\centering
		\includegraphics[scale=0.6]{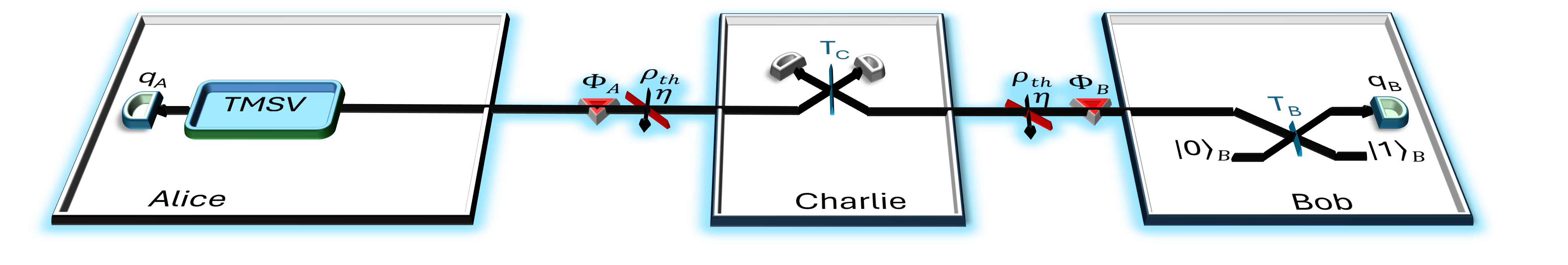}
\caption{CV-QKD protocol for overcoming the PLOB bound along with the phase noise.
Phase noises ($\phi_A$) and ($\phi_B$) act on the modes sent by Alice and Bob to Charlie’s station, respectively, while thermal noise ($\rho_{th}(\eta)$) affects both transmitted modes. $T_{C}$ and $T_{B}$ denote $(50{:}50)$ beamsplitters. $(\ket{0}_B)$ represents the auxiliary vacuum mode at Bob’s station, and $(\ket{1}_B)$ denotes a single-photon ancilla state at Bob’s station. $(q_A)$ and $(q_B)$ represent the heterodyne measurements performed by Alice and Bob, respectively. In Charlie's station, the detectors are photon-number-resolving detectors (PNRDs).
Alice selects one mode from the TMSV state and transmits it to Charlie through a channel affected by both thermal noise and phase noise. Simultaneously, Bob generates a single-photon entangled state and directs one of its modes toward Charlie as well. Upon receiving the modes, Charlie interferes them using a $50{:}50$ beamsplitter and performs photon-number-resolving detection. A single detection event (click) signifies the presence of strong correlations between Alice and Bob. Subsequently, Bob measures his remaining mode via heterodyne detection. The gain of the NLA is adjusted according to Charlie’s precise position between Alice and Bob, as well as the transmissivities of the beamsplitters, $T_{C}$ and $T_{B}$.
\label{fig:CirNLAPh}}
\end{figure*}

\begin{figure*}[!htb]
	\centering
		\includegraphics[scale=0.6]{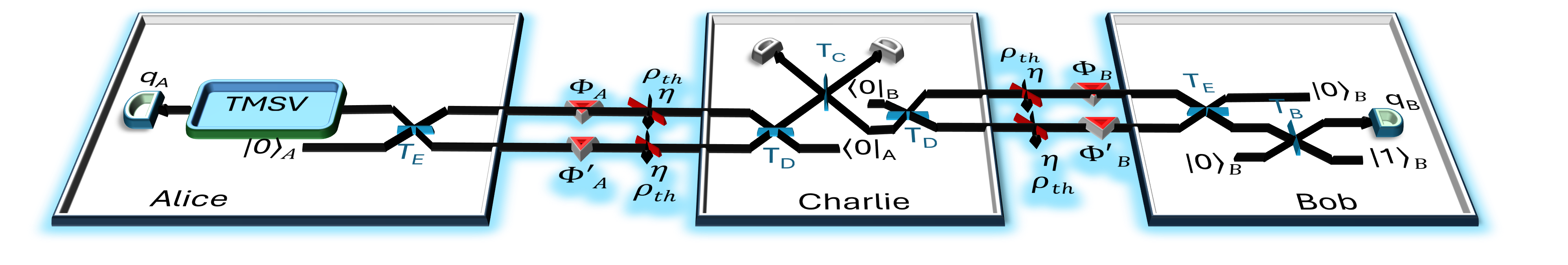}
\caption{CV-QKD protocol for overcoming the PLOB bound with UA scheme. 
Phase noises ($\phi_A$ and $\phi'_A$) and ($\phi_B$ and $\phi'_B$) act on the modes sent by Alice and Bob, respectively, after encoding and before reaching Charlie’s station, while thermal noise ($\rho_{th}(\eta)$) affects all transmitted modes. $T_{E}$, $T_{D}$, $T_{C}$ and $T_{B}$ denote $50{:}50$ beamsplitters. $\ket{0}_A$ and $\ket{0}_B$  represent auxiliary vacuum modes at Alice’s and Bob’s stations, respectively. $\bra{0}_A$ and $\bra{0}_B$ denote vacuum detections at Charlie’s station, while $(\ket{1}_B)$ denotes a single-photon ancilla state at Bob’s station. $q_A$ and $q_B$ represent the heterodyne measurements performed by Alice and Bob, respectively. The detectors at Charlie’s station are photon-number-resolving detectors (PNRDs).
Alice prepares a TMSV state along with a vacuum state. After completing the encoding step using a $50{:}50$ beamsplitter, she sends both modes to Charlie’s station. In parallel, Bob prepares a single-photon entangled state together with a vacuum state. Similarly, after performing the encoding step, Bob sends both modes to Charlie’s station. At Charlie’s end, the decoding step is carried out using a $50{:}50$ beamsplitter for both Alice and Bob. Upon successful heralding on the vacuum outcomes for the respective vacuum modes of Alice and Bob, Charlie interferes the remaining modes using a $50{:}50$ beamsplitter and performs photon-number-resolving detection.
\label{fig:CIRNLAUA}}
\end{figure*}

We begin by reviewing the no switching CV-QKD protocol used throughout this work, focusing on its entanglement-based version, which allows for a more straightforward theoretical study \cite{grosshans2003virtual}. In this framework, Alice initially creates a two-mode squeezed vacuum (TMSV) state
$\ket{\text{TMSV}} = (\cosh{r})^{-1} \sum_{n = 0}^{\infty}(-\tanh{r})^{n}\ket{n, n}$
This is a two-mode Gaussian state \cite{BAC19, weedbrook2012gaussian} which can be fully characterized by its covariance matrix (CM) (see Appendix \ref{BRIEF REVIEW OF THE PHASE-SPACE FORMALISM} for additional details):

\begin{align}
   \Sigma_{\text{TMSV}} = \begin{pmatrix}
        \mathbb{A} \mathbf{I}_{2} & \mathbb{C} \mathbf{\sigma}_{2} \\
        \mathbb{C} \mathbf{\sigma}_{2} & \mathbb{B}\mathbf{I}_{2} 
    \end{pmatrix},
\end{align}
where $\mathbb{A} =\mathbb{B} = \cosh{2r}$ corresponding to the input modulation variance of the protocol, is the variance, $\mathbb{C} = \sinh{2r}$ is the covariance, $ \mathbf{I}_{2} = \text{diag}(1, 1)$, and $\mathbf{\sigma}_{2}$ represents the Pauli $z$ matrix. All variables are expressed in shot-noise units and is normalised such that the vacuum covariance matrix is simply the identity.

Alice performs a heterodyne measurement on the first mode, while the second mode is transmitted to Bob through an untrusted quantum channel, which is often modelled as a thermal-loss channel. In this case the channel is characterized by a transmissivity given by $\eta = 10^{-L/10 D}$, where $D$ denotes the transmission distance in kilometers, and $L \thicksim 0.2$ dB/km represents the typical loss coefficient for optical fibres operating at 1550 nm \cite{li2005optical,larsen2019fiber}.  Furthermore, the presence of excess noise $\epsilon$, which may arise due to various imperfections or the action of an eavesdropper, is modelled by a single-mode thermal bath containing $n_{\text{th}} = \eta \epsilon / 2(1 - \eta)$ photons \cite{lodewyck2007quantum}. The losses and imperfections impact the signal received by Bob, who will detect an added noise $\varkappa  = (1 - \eta)/\eta + \epsilon$, resulting in an overall thermal-loss channel. Consequently, the quantum state shared between Alice and Bob remains Gaussian, characterized by the covariance matrix (CM) \cite{laudenbach2018continuous, ferraro2005gaussian, serafini2023quantum}.

\begin{align}
   \Sigma^{\text{(Th)}}_{\text{AB}} &=  \begin{pmatrix}
       \Sigma_{\text{A}} & \Sigma_{\text{C}} \\
       \Sigma_{\text{C}} & \Sigma_{\text{B}}
    \end{pmatrix}, \\
    & =  \begin{pmatrix}
        \mathbb{A} \mathbf{I}_{2} & \sqrt{\eta}\mathbb{C} \mathbf{\sigma}_{2} \\
        \sqrt{\eta}\mathbb{C} \mathbf{\sigma}_{2} & \eta (\mathbb{A} + \varkappa )\mathbf{I}_{2} 
    \end{pmatrix}.  \label{COVth}
\end{align}
Upon receiving the signal, Bob measures his mode
with heterodyne detection \cite{weedbrook2004quantum}.

All the necessary information to perform the security analysis is contained in the covariance matrix (\ref{COVth}). According to the Gaussian formalism \cite{serafini2023quantum, olivares2012quantum, weedbrook2004quantum}, when Alice and Bob perform detection on their own signals they get a bivariate Gaussian distribution with zero mean and covariance $\Sigma_{\text{B}} + \Lambda^{\text{(m)}}_{\text{B}} $, where $\Lambda^{\text{(m)}}_{\text{A}} = \Lambda^{\text{(m)}}_{\text{B}} = \mathbb{I}_{2}$ where,
$\Lambda^{\text{(m)}}_{\text{A (B)}}$ is the CM with the heterodyne detection.
The mutual information between Alice and Bob is then given by
\begin{align}
    I_{\text{AB}} = & H(A) + H(B) - H(AB) \\
    = & \log_{2} \sqrt{\frac{\det(\Sigma_{\text{A}} + \Lambda^{\text{(m)}}_{\text{A}})\det(\Sigma_{\text{B}} + \Lambda^{\text{(m)}}_{\text{B}})}{\det(\Sigma_{\text{AB}} + (\Lambda^{\text{(m)}}_{\text{A}} \bigoplus  \Lambda^{\text{(m)}}_{\text{B}}))}},
\end{align}
where $H(x) = -\int p(x)\log_{2}p(x) \,\textrm{d}x$ or $H(x) = -\sum_{x}p(x)\log_{2}p(x) $ is the Shannon entropy
of $p(x)$.

Throughout this paper we concentrate on a reverse reconciliation scheme, which has been shown to provide stronger security than direct reconciliation \cite{grosshans2003quantum,pirandola2020advances}. We also assume the eavesdropper (Eve) can perform collective attacks, which constitute the most powerful attacks available to them in the asymptotic limit of an infinite data set \cite{grosshans2005collective}.
The SKR in the asymptotic limit is
\begin{align}
    \kappa = \beta I_{\text{AB}} - \chi_{\text{BE}}, \: 0\leq \beta \leq 1 \label{skr}
\end{align}
where $\beta$ is the reconciliation efficiency and the Holevo information, $\chi_{\text{BE}}$, quantifies the amount of information that Eve can extract \cite{holevo2002capacity}.
This can be can be upper bounded solely from the covariance matrix as
\begin{equation}
\chi_{BE} = S(AB) - S(A|B).
\tag{B6}
\end{equation}
where $S(AB)$ and $S(A|B)$ are determined from the symplectic eigenvalues $g_k$
of the corresponding covariance matrices $\Sigma_{AB}$ and $\Sigma_{A|B}$.
They are computed using
\begin{equation}
S(\cdot) = \sum_{k=1}^{N}
\left[
\frac{g_k + 1}{2}\log_2\!\left(\frac{g_k + 1}{2}\right)
-
\frac{g_k - 1}{2}\log_2\!\left(\frac{g_k - 1}{2}\right)
\right],
\end{equation}
where $N$ represents the number of mode.
The above estimation is derived under the assumption of infinitely large modulation and detection data. In practical CV-QKD systems, however, the data length is necessarily finite, which introduces statistical fluctuations in the estimated parameters. Despite these finite-size effects, the asymptotic key rate remains a benchmark for validating the feasibility of the proposed scheme and for guiding experimental design.

In the next section, we investigate how the simultaneous presence of phase noise, pure loss and thermal noise affect the channel (see Fig.~\ref{fig:CirPh} for set-up), and we analyse the resulting key rate, $\kappa$, as a function of the transmission distance under these noise conditions. We then study the behaviour of $\kappa$ as a function of the transmission distance for the UA-assisted protocol (Fig.~\ref{fig:CIRUA}) in Section~\ref{UA-Assisted CV QKD}. Subsequently, we incorporate phase noise into the relay scheme once again (Fig.~\ref{fig:CirNLAPh}). Finally, we apply the hybrid UA–NLA protocol and study the behaviour of $\kappa$ as a function of the transmission distance under these noise conditions, following the scheme described in Section~\ref{Hybrid UA-QS CV QKD}.

Throughout the analysis, we assume a fixed reconciliation efficiency of $\beta \sim 0.95$ and an excess noise $\epsilon = 0.02$ \cite{leverrier2009unconditional}. The excess noise is taken to originate from the environment and is modelled as input thermal noise.

\section{UA-Assisted CV QKD}
\label{UA-Assisted CV QKD}

\begin{figure*}[!htb]
	\centering
		\begin{subfigure}{\columnwidth}
             \caption{ \label{plot:PHUA0.1}}
			\includegraphics[width=1.0\columnwidth]{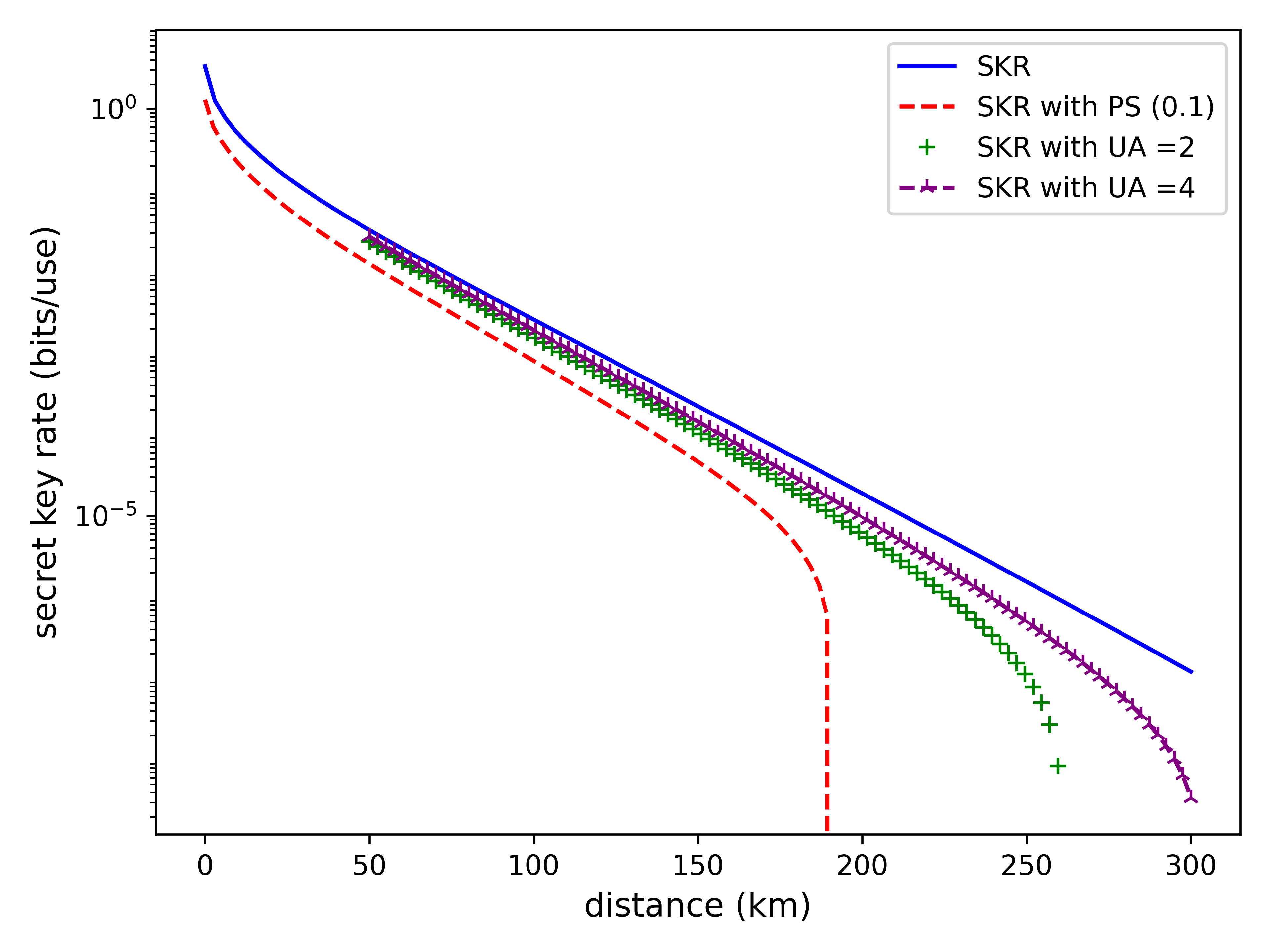}
			
		\end{subfigure}
  		\begin{subfigure}{\columnwidth}
              \caption{ \label{plot:PHUA0.3}}
			\includegraphics[width=\columnwidth]{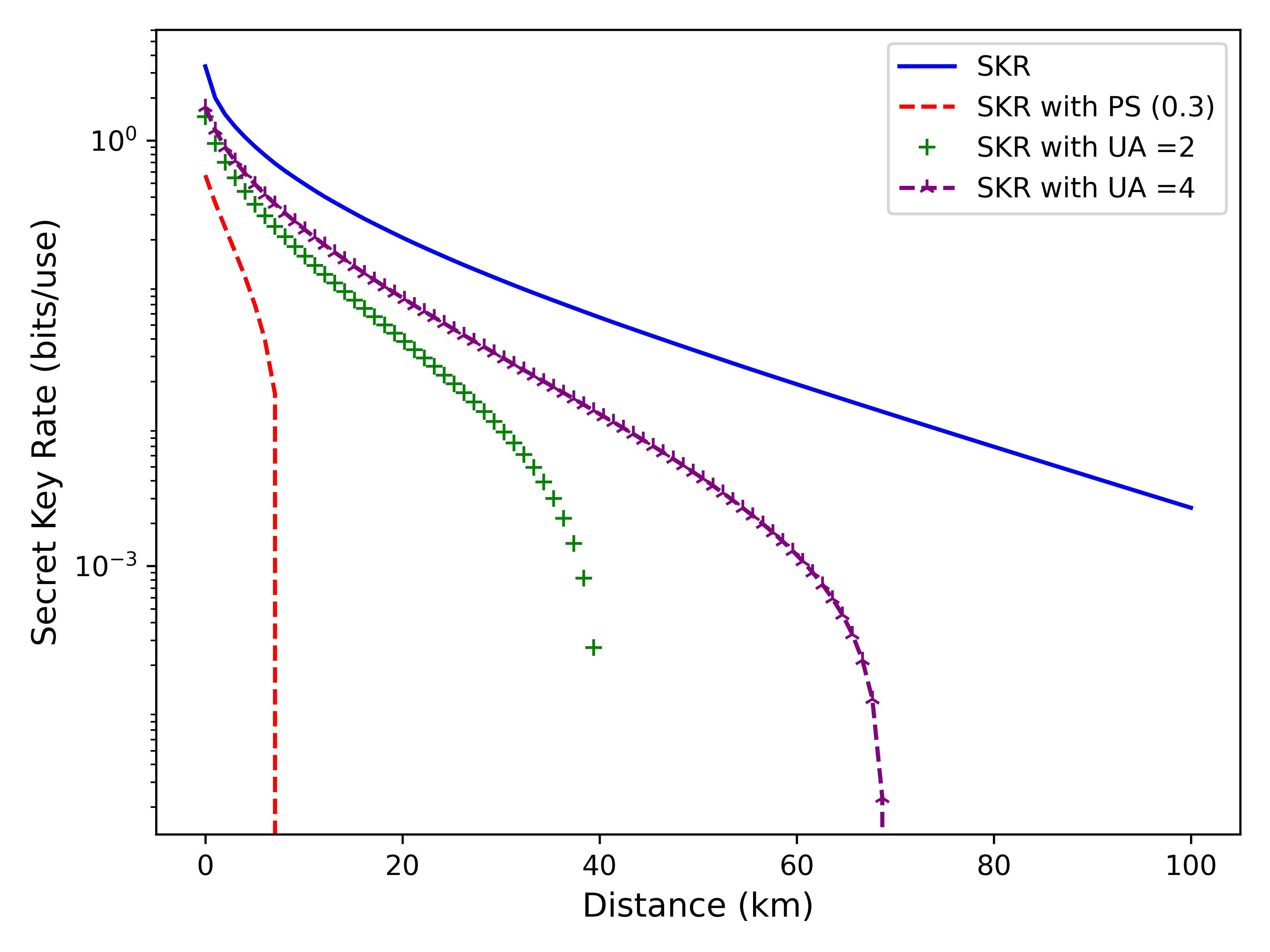}
		\end{subfigure}
		\caption{Secret key rate as a function of transmission distance with $\beta = 0.95$ and excess noise $\epsilon = 0.02$. The solid blue line represents the case without phase noise.
        \ref{plot:PHUA0.1} and  \ref{plot:PHUA0.3} represent SKR with phase noise $\sigma = 0.1$ and $\sigma = 0.3$ respectively. In the both the plots, the dashed, red line represents SKR with phase noise. The Green line with marker `$+$' and purple dashed-$+$ line clearly shows the improvement after applying the UA protocol once ($N=2$) and twice ($N=4$) respectively.
  \label{fig:CIRplotsua}}
\end{figure*}

\begin{figure*}[!htb]
	\centering
		\begin{subfigure}{\columnwidth}
             \caption{ \label{plot:PHNLA0.1}}
			\includegraphics[width=1.0\columnwidth]{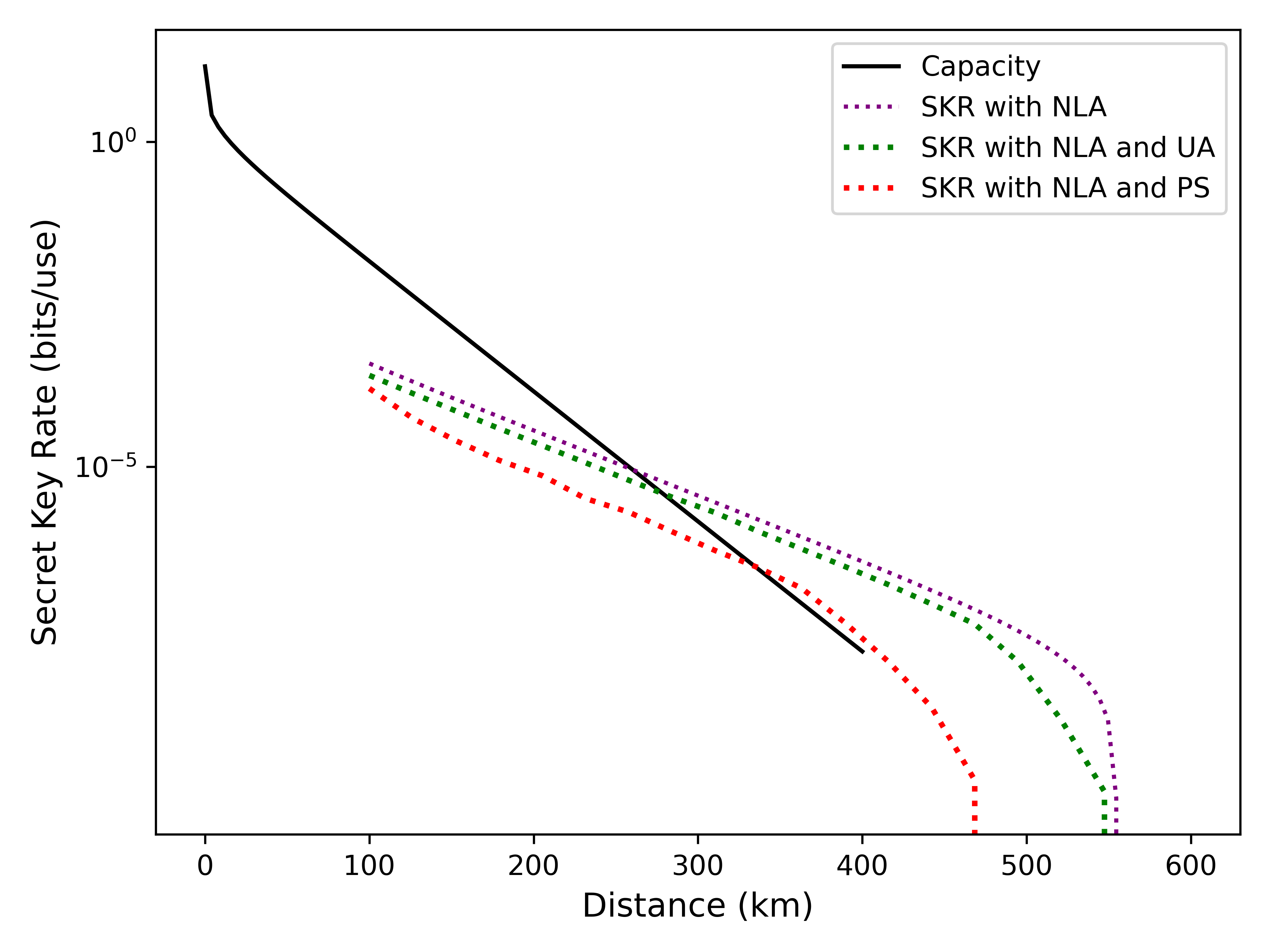}
			
		\end{subfigure}
  		\begin{subfigure}{\columnwidth}
              \caption{ \label{plot:PHNLA0.5}}
			\includegraphics[width=1.0\columnwidth]{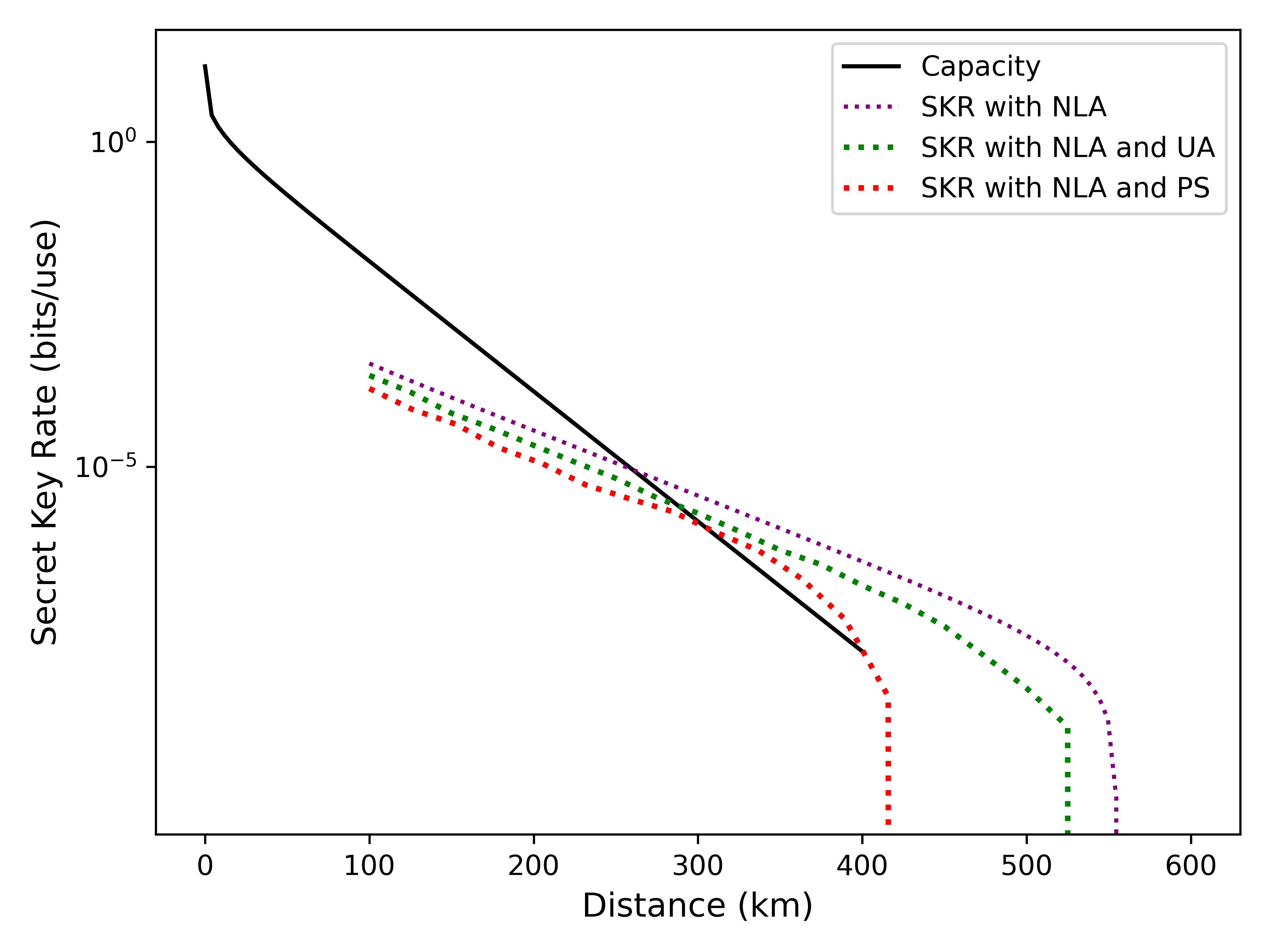}
		\end{subfigure}
		\caption{SKRs as functions of the distance in kilometers with $\beta = 0.95$ and excess noise $\epsilon = 0.02$ with and without applying hybrid UA-NLA protocol. Capacity is the PLOB bound. \ref{plot:PHNLA0.1} and  \ref{plot:PHNLA0.5} represent SKR with phase noise $\sigma = 0.1$ and $\sigma = 0.5$ respectively. In both the plots, red dotted line represents SKR with phase noise when employing the distributed NLA protocol. The green dotted line clearly shows the improvement after applying the hybrid UA-NLA protocol. The purple dotted line in both the plots show the keyrate with an NLA but no phase noise. 
  \label{fig:CIRplotsNLA}}
\end{figure*}

In this section, we consider a phase-noise and a thermal-loss channel, as illustrated in Fig. \ref{fig:CirPh}. The phase noise in this setup is modelled as a stochastic phase shifter (PS), with the phase drawn from independent Gaussian distributions for each mode. Physically, this could arise from small, independent fluctuations in path distance with respect to the local oscillator. For simplicity we have assumed the phase noise is independent of propagation distance (which is unlikely to be true in practice). We also present the UA protocol, as illustrated in Fig.~\ref{fig:CIRUA}. In this protocol, Alice, in addition to preparing the TMSV state, also generates an auxiliary mode in the vacuum state. Subsequently, Alice combines the second mode of the TMSV state with the vacuum state using an encoding beamsplitter ($T_{E}$ in Fig.~\ref{fig:CIRUA}), which is a $50{:}50$ beamsplitter.
This two mode signal then propagates through the thermal loss and phase noise channel before being recombined by Bob using the decoding beamsplitter ($T_{D}$ in Fig.~\ref{fig:CIRUA}). It is assumed that the phase noises affecting the two-modes are of equal magnitude, but are independent.
Finally, Bob performs vacuum detection on the secondary received mode.
After performing a Gaussian measurement on the other mode, Bob obtains the phase-reduced state.

As discussed in \cite{swain2024improving}, the resulting output state is approximately Gaussian. Consequently, the protocol depicted in Figs. (\ref{fig:CirPh}, \ref{fig:CIRUA}), after vacuum heralding, results in a similar output state to the scheme described in Section \ref{THE ORIGINAL QKD PROTOCOL}.
Thus, in our approach, it is sufficient to compute the covariance matrices (CMs) $ \Sigma^{(\text{PS})}_{AB} $ and $ \Sigma^{(\text{UA})}_{AB} $, corresponding to the states $\rho_\text{PS}$ and $\rho_\text{UA}$, respectively, in order to carry out the security analysis. 
These CMs can be written
\begin{align}
   \Sigma^{(\text{PS})}_{AB} &= \begin{pmatrix}
        \mathbb{A}_{\text{PS}} \mathbf{I}_{2} & \mathbb{C}_{\text{PS}} \mathbf{\sigma}_{2} \\
        \mathbb{C}_{\text{PS}} \mathbf{\sigma}_{2} & \mathbb{B}_{\text{PS}}\mathbf{I}_{2} 
    \end{pmatrix} \\
     & = \begin{pmatrix}
        \cosh{2r} \mathbf{I}_{2} & \sqrt{\eta} \sinh{2r} \cos{\phi}\mathbf{\sigma}_{2} \\
        \sqrt{\eta} \sinh{2r} \cos{\phi} \mathbf{\sigma}_{2} & \eta(\cosh{2r} + \frac{1- \eta}{\eta} + \epsilon)\mathbf{I}_{2} \label{COVpsfinal}
    \end{pmatrix}\\
\Sigma^{(\text{UA})}_{a1b1} &= \begin{pmatrix}
        \mathbb{A}_{\text{UA}} \mathbf{I}_{2} & \mathbb{C}_{\text{UA}} \mathbf{\sigma}_{2} \\
        \mathbb{C}_{\text{UA}} \mathbf{\sigma}_{2} & \mathbb{B}_{\text{UA}}\mathbf{I}_{2} 
    \end{pmatrix}
\end{align}
where
\begin{align}
\mathbf{Z} =  & \big(\frac{e^{i \phi_{b1}} + e^{i \phi_{a2}} }{2}\big); \: \phi_{j} \sim \mathcal{N}(0, \sigma^2) \\
    \mathbb{A}_{\text{UA}} \equiv & \cosh{2r}\sqrt{\Big(\Re(\mathbf{Z})\Big)^{2} + \Big(\Im(\mathbf{Z})\Big)^{2}} \\
    \mathbb{B}_{\text{UA}} \equiv & \eta \Big(\mathbb{A}_{\text{UA}} + \frac{1- \eta}{\eta} + \epsilon\Big) \\
    \mathbb{C}_{\text{UA}} \equiv & \sqrt{\eta}\sinh{2r}\sqrt{\Big(\Re(\mathbf{Z})\Big)^{2} + \Big(\Im(\mathbf{Z})\Big)^{2}} \nonumber \\ 
   & \cos{\Big(\arctan{\frac{\Im(\mathbf{Z})}{\Re(\mathbf{Z})}}\Big)}
\end{align}
where $\Re(\mathbf{Z})$ and $\Im(\mathbf{Z})$ is the real and imaginary part of $\mathbf{Z}$.
The step by step calculations for $\Sigma^{(\text{PS})}_{AB}$ and $\Sigma^{(\text{UA})}_{a1b1}$, including the explanation of the Kraus operator representation, are provided in Appendix \ref{CALCULATION FROM VARIOUS PROTOCOL}.  These results demonstrate how the presence of phase noise acts do reduce the covariance, $\mathbb{C}$, terms, while the adition of UA reduces this impact, but also modifies both the variance terms  $\mathbb{A}$ and  $\mathbb{B}$.
We evaluate the mutual information and the Holevo information following the procedure outlined in Section \ref{THE ORIGINAL QKD PROTOCOL}, substituting $\rho_{\text{AB}} \to \rho^{p}_{\text{AB}}$ with $p = (\text{PS}, \text{UA})$, and then use Eq. \ref{skr} to compute the corresponding SKRs.

\begin{align}
   \kappa^{(\text{PS})} &= \beta I^{(\text{PS})}_{\text{AB}} - \chi^{(\text{PS})}_{\text{BE}} \\
   \kappa^{(\text{UA})} &= P^{(\text{UA})}\big(\beta I^{(\text{UA})}_{\text{AB}} - \chi^{(\text{UA})}_{\text{BE}}\big)
\end{align}
where $\kappa^{(\text{UA})}$ is defined as the standard raw key rate multiplied by the success probability of the UA protocol, denoted by $P^{(\text{UA})}$.
 Both $\kappa^{(\text{PS})}$ and $\kappa^{(\text{UA})}$ correspond to the asymptotic secret key rates without and with the application of the UA protocol, respectively.
A detailed explanation of the UA protocol is provided in Appendix \ref{Unitary Averaging in a nutshell}. The procedure for applying the UA protocol once is given in Appendix \ref{AliceUA}. To apply the protocol twice, the input state is combined with two ancilla modes using additional $50{:}50$ beamsplitters as per the encoding step. Each mode then undergoes identical and independent phase noise. The decoding beamsplitter network subsequently reverses the encoding. After heralding vacuum in the two auxiliary modes, the resulting output state exhibits improved quality.
The corresponding calculations for the double application of the UA protocol can be carried out by following the methods presented in Appendices \ref{Unitary Averaging in a nutshell} and \ref{AliceUA}.

Next, we compare the SKRs obtained with and without applying the UA protocol. The red dashed lines in Fig.~\ref{fig:CIRplotsua} illustrate that phase noise significantly limits the transmission distance compared with the solid black line, which represents the case without phase noise.
In Fig. \ref{plot:PHUA0.1}, for a phase-noise channel with $\sigma = 0.1$ and a thermal-loss channel characterized by an excess noise of $\epsilon = 0.02$, the SKR without UA drops to zero around 190~km. However, with the application of UA, the key rate extends up to approximately 300~km, indicating an improvement of about 110~km.
Similarly, in Fig. \ref{plot:PHUA0.3}, for a higher phase noise of $\sigma = 0.3$, the asymptotic key rate without UA vanishes around 5~km. In contrast, applying the UA protocol twice extends the key rate up to nearly 70~km.

Overall, the UA protocol demonstrates significantly enhanced performance in a decent phase-noise regime. In the following, we first discuss the NLA protocol. We then describe the how the NLA protocol is effected by the presence of phase noise. Finally, we combine the UA and NLA protocols to investigate the resulting secret key rates under conditions of phase noise.

\section{Hybrid UA-NLA CVQKD}
\label{Hybrid UA-QS CV QKD}

Noiseless-linear amplifiers are an idealised quantum amplifier of gain $g$ which takes the coherent state $\left\langle\alpha\right\rangle\rightarrow\left\langle g\alpha\right\rangle$ and is thus poportional to the operator $g^{\hat{n}}$. Generalised quantum scissors \cite{ralph2008nondeterministic, pegg1998optical, guanzon2022ideal} accurately performs this this noiseless linear amplification (or deamplification) of arbitrary input states, only up to a specified Fock number (k). The operation involves combining the input state with an entanglement resource state and post-selecting on events where (k) detectors each register a single click. This process effectively teleports and amplifies the input state onto the output mode. If the measurement and output mode are in different places, it becomes a teleamplifier. 
In the $k=1$ case, the amplification gain is determined by the transmissivity of Bob’s beam splitter and this is often simply refered to as a quantum scissor.
When the measurement is positioned midway between Alice and Bob, the success probability of the teleamplifier scales as the square root of the total channel transmissivity. Consequently, the protocol can surpass the PLOB bound, which itself scales linearly with the total channel transmissivity; a result previously reported in \cite{winnel2021overcoming, guanzon2022ideal}.

We consider a scenario analogous to Fig. (1b) in Ref. \cite{winnel2021overcoming}, now with the addition of phase noise (Fig. \ref{fig:CirNLAPh}). Alice prepares an EPR state, performs heterodyne detection, and then sends one of her modes to Charlie through a channel affected by both thermal noise and phase noise. In parallel, Bob prepares a single-photon entangled state and transmits one of its modes to Charlie. At Charlie’s station, the incoming modes are interfered and measured using photon-number-resolving detectors. A single-photon detection event heralds strong correlations between Alice and Bob. Bob then heterodynes his remaining mode. The overall success probability scales with the square root of the channel transmissivity, indicating that the teleamplifier enhances the rate–distance scaling. The red dotted curves in Fig. \ref{fig:CIRplotsNLA} illustrate the impact of phase noise on the distributed NLA protocol.

We introduce a hybrid approach that combines the UA and distributed NLA protocols to enable long-distance CV-QKD, simultaneously mitigating phase noise and loss, allowing a positive key rate at distances surpassing the PLOB bound.
In this scheme, Alice prepares a TMSV state along with an auxiliary vacuum state. After the encoding step, she transmits both entangled modes to Charlie through a thermal-noise channel and a phase-noise channel. In parallel, Bob prepares a single-photon entangled state accompanied by a vacuum state. Following his own encoding step, Bob sends the encoded modes to Charlie as well.  Alice retains mode $a_1$, while Bob keeps mode $b_2$. To facilitate the calculations presented in Appendix~\ref{CALCULATION FROM VARIOUS PROTOCOL}, we denote the modes more generally by $a_i$ and $b_i$, with $i = 1,2,3$.

At Charlie’s station, the decoding operation and vacuum detection for both Alice’s and Bob’s modes are performed. The modes $(a_2 \;\text{and}\; b_1 )$ are then interfered, and photon-number-resolving detection is carried out. A single detection click heralds enhanced correlations between Alice and Bob. The success probability scales with the square root of the overall channel transmissivity, making the teleamplifier effective in enhancing the rate–distance scaling. The gain of the NLA is determined by Charlie’s position between Alice and Bob, as well as by the transmissivities of the beam splitters. A schematic illustration of the setup is provided in Figs. (\ref{fig:CirNLAPh}, \ref{fig:CIRNLAUA}).
Comprehensive analytical derivations and numerical simulations supporting this scheme are presented in Appendix~\ref{EMPLOYING QUANTUM SCISSOR AND UNITARY AVERAGING}. The corresponding secret key rates (SKRs) are computed using Eq.~\ref{skr}.
\begin{equation}
    \kappa^{(\text{QS})} = P^{(\text{QS})}P^{(\text{UA})}\left(\beta I^{(\text{QS})}_{\text{AB}} - \chi^{(\text{QS})}_{\text{BE}}\right) \label{PS1}
\end{equation}
With the absence of UA, SKR is,
\begin{equation}
    \kappa^{(\text{QS})} = P^{(\text{QS})}\left(\beta I^{(\text{QS})}_{\text{AB}} - \chi^{(\text{QS})}_{\text{BE}}\right) \label{PS2}
\end{equation}
where $P^{(\text{QS})}$ is the total success probability where $P^{(\text{QS})} = 2  \Tr\big((\mathbf{I}_{a1} \otimes \bra{0} \otimes \bra{1}\otimes \mathbf{I}_{b2} )\rho_{\text{AB}} \big) $.
The factor of $2$ arises because there are two successful click patterns; one pattern imparts a passive $\pi-$phase shift on the output state, which is straightforward to correct once Charlie informs Bob which detector clicked. The success probability of the UA protocol is denoted by $P^{(\text{UA})}$.
The PLOB bound is also shown which is $ -\log_{2}(1- \eta_{\text{PLOB}})$ \cite{pirandola2017fundamental} where $\eta_{\text{PLOB}}= 10^{-L/10 D} $, $D$ is the distance in km. 

Figures \ref{fig:CIRplotsua} and \ref{fig:CIRplotsNLA} demonstrate the improvements to key rates when utilising UA without and with the distributed NLA repeater respectively. They demonstrate that even  with and without phase noise and demonstrating the advantage of UA. This shows that even for moderate phase noise ($\sigma = 0.1$ for Figures \ref{plot:PHUA0.1} and \ref{plot:PHNLA0.1}) a notible improvment in key rate is achievable when utilising UA, with the maximum transmission distance improving from $\sim190$ km to $>500$ km when utilising the full UA-NLA protocol. More dramatic improvements can be seen in the higher phase noise regime, where similar transmission distances can be achieved for $\sigma = 0.5$ with the new protocol despite a maximum transmission distance of $<10$ km for $\sigma = 0.3$ for the standard no switching protocol. This protocol is tolerant to excess noise, phase noise and loss, and suggests it could improve the scaling of many recent CV QKD protocols. Of particular note is that it proves to be particularly powerful in the presence of a high phase noise channel.

It is important to note that the overall SKR is determined by the product of the raw key generation rate and the corresponding success probabilities (Eqs. \ref{PS1} and \ref{PS2}). As a result, the success probability scales proportionally with the final SKR.   Consequently, rather than plotting the success probability as a separate metric, its impact is inherently embedded within the SKR curves presented in Fig \ref{fig:CIRplotsua} and \ref{fig:CIRplotsNLA}. Any reduction in the success probability, whether due to higher UA or accumulated transmission losses, directly manifests as a proportional decrease in the plotted SKR, accurately reflecting the practical performance trade-offs of the system.

\section{Experimental Considerations}

In practical implementations of continuous-variable quantum protocols, transmitting signals through a common transmission line, such as a standard optical fiber, inevitably subjects the light to macroscopic phase fluctuations and loss. Environmental factors, including thermal drift and acoustic vibrations, cause slow, global changes in the fiber's refractive index and physical length. 
To successfully operate over such channels, the system must actively track and compensate for this global drift. 
This is typically achieved by co-propagating a bright, classical reference laser in an orthogonal mode (such as a different polarisation or a different temporal mode) alongside the quantum signal. Because this bright reference travels through a similar physical medium at the same time, it acquires the same macroscopic phase shift. At the receiver the classical beam is separated from the quantum signal and used as a local reference that can be used for phase sensitive measurements and to lock phases between different quantum signals.

The phase noise variables utilised in the models of this paper do not represent the macroscopic drifts of the transmission line. Instead, they represent the phase noise that remains after the active phase-reference compensation has been applied.
This phase noise is dominated by fundamentally uncorrelated and stochastic sources arising from small differences in the medium seen by the quantum signal and the bright reference as well as imperfections in the active compensation. 
It is reasonable to assume that this residual phase noise is independent and identically distributed (i.i.d.) across different transmission channels.

In the set-up shown in Fig.\ref{fig:CIRNLAUA} classical references would co-propagate along the two channels from Alice to Charlie and the two channels from Bob to Charlie. By comparing the classical references in the different channels they can be used to: (i) balance the UA interferometer between Alice and Charlie; (ii) balance the UA interferometer between Bob and Charlie; and (iii) lock the phase between the inputs to the $T_C$ beamsplitter. The protocol will then act to suppress the residual phase noises ($\phi_A, \phi_A', \phi_B, \phi_B'$) and the loss, as described in the text. We note that the combination of NLA with CV systems (albeit in simpler situations) has been experimentally demonstrated \cite{ferreyrol2010implementation, ulanov2015undoing}.

\section{Conclusion}
\label{Conclusion}

In this manuscript, we have investigated the use of UA to enable long-distance CV QKD under realistic conditions, and in particular demonstrated its power when combined with the distributed NLA repeater. These benefits are maintained despite non-unit reconciliation efficiency, non-zero excess channel noise, and non-zero phase noise.

First, we introduced phase noise together with thermal noise in the squeezed-state protocol, as illustrated in Fig. 1. The degradation due to phase noise is clearly demonstrated by the red dotted line in Fig. 5. Next, we apply the UA protocol, shown in Fig. 2, to mitigate the phase error. The resulting improvement is illustrated by the green line with ``+” markers and the purple dashed line with ``+” markers, which indicate a clear enhancement compared to the noisy case.
In the next part of the paper, we considered the scenario shown in Fig. 3, where phase noise is again introduced alongside thermal noise. In this case, the NLA is used to correct the thermal noise. When applied to CV-QKD at Charlie’s station, this approach enables the protocol to surpass the PLOB bound without requiring quantum memories. The impact of phase noise in this scenario is shown by the red dotted line in Fig. 6.
Finally, in Fig. 4, we propose combining the UA scheme (previously used to mitigate phase noise) with NLA to correct thermal noise. When implemented in the CV-QKD protocol at Charlie’s station, this combined approach also surpasses the PLOB bound without requiring quantum memories. The resulting improvement can be observed in Fig. 6, indicated by the green dotted line.

In brief, We propose a practical CV quantum communication protocol based on a unitary averaging techniques previously employed to mitigate phase noise and NLA, which, when applied to CV QKD in the Charlie's station, surpasses the PLOB bound without requiring quantum memories, even in the presence of significant phase noise. The normalisation constant of the output state after the vacuum detection yeilds the success probability of UA, which was found to be sufficiently high with the noise channels considered. When the station is placed symmetrically between Alice and Bob with balanced beamsplitters, the NLA success probability scales as $\sqrt{n}$, and the NLA gain across the second link compensates for the loss in the first link. Consequently, the UA-NLA protocol protects against both phase noise and loss, and is robust to thermal noise.

These results offer new insights into the practical use of hybrid UA-NLA under realistic conditions, both for one-way communication and for end-to-end transmission over quantum repeater chains \cite{dias2020quantum, pirandola2019end, zhong2021proof}.

\section{Acknowledgement}
SNS appreciates the discussions with Laura Serino regarding \textit{Experimental Implementation Strategies}.
SNS was supported by the Sydney Quantum Academy, Sydney, NSW, Australia. This work was
partially supported by the Australian Research Council Centre of Excellence for Quantum Computation and
Communication Technology (Project No.CE170100012).

\appendix
\section{Unitary Averaging in a nutshell}
\label{Unitary Averaging in a nutshell}
\begin{figure}[!htb]
	\centering
		 \includegraphics[width=0.6\linewidth]{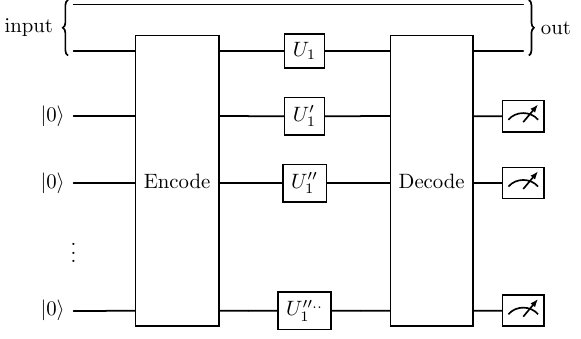}
		\caption{  A passive unitary averaging scheme utilises a beam splitter network for redundant encoding over $n$ modes. One mode of a two-mode state is evenly distributed across $n$ transmission modes. Each mode undergoes independent single-mode unitary noise. The decoding network reverses the encoding process, and upon heralding $n-1$ error modes in the vacuum state, the output state exhibits reduced noise.
  \label{CVUAcir}}
\end{figure}

We begin by reviewing previous work on Gaussian systems, in which the impact of UA on a continuous-variable (CV) system is modelled using a two-mode squeezed vacuum state, as described in \cite{swain2024improving}.
In this model, a passive beam splitter encoding network is employed to place the input state into an equal superposition of $n$ spatial modes. Each of these modes undergo independent transformations, previously simply transmission though phase noise, we are now considering arbitrary linear optical unitary transformations. The decoding network then inverts the encoding such that each superposition should constructively interfere in the top most modes. The remaining `error modes' are then heralded in the vacuum state, such that the heralded output state exhibits suppressed noise. The behaviour is characterised by the resulting output squeezing, purity, and entanglement. 
The two-mode squeezed vacuum state was chosen for characterising the system because it can effectively represent various input states through projective measurement. 
Fig.\ref{CVUAcir} illustrates the circuit discussed in Ref. \cite{swain2024improving}. Below, we provide a brief summary of the corresponding results.

The output state from the the protocol is,

    \begin{align}
    \ket{\psi_{\text{out}}} &= \frac{1}{\mathbf{N}} \left(\cosh{r}\right)^{-1}\sum_{N} (-1)^{N} [e^{i \phi_{\beta}} \tanh{r'} ]^{N} \ket{N, N}\nonumber\\
    &=\hat{S}\left(\chi'\right)\ket{0}\ket{0}
\end{align}
where $\chi'=r'e^{i\phi_{\beta}}$ and the phase terms $\phi_j$ vary randomly and independently around some mean value. We define $\alpha e^{i\phi_{\beta}}= \frac{e^{ i \phi_{1}} + e^{ i \phi_{2}} +...+e^{ i \phi_{n}}}{n}$ and $\tanh{r'} = \alpha\tanh{r}$ with normalisation constant $\mathbf{N}  = \frac{\cosh{r'}}{\cosh{r}}$ and 
probability of success  $P = |\mathbf{N}|^{2}$. 

In the covariance matrix form we have,
\begin{equation}
     \Sigma_{\text{out}(4 \cross 4)} = \left\langle\begin{pmatrix}
          A  &  C  \\
          C^{T} &  B
     \end{pmatrix}\right\rangle \label{eq:covariance matrix}
 \end{equation}
where,
\begin{align}
    A = B 
    & = \begin{pmatrix}
        \frac{1+ \tanh{r^{'}}^{2}}{1 - \tanh{r^{'}}^{2}} & 0\\
        0 &  \frac{1+ \tanh{r^{'}}^{2}}{1 - \tanh{r^{'}}^{2}}
    \end{pmatrix} \label{aeq}
\end{align}

\begin{align}
    C   
    & = \begin{pmatrix}
       - \frac{2\tanh{r^{'}}}{1 - \tanh{r^{'}}^{2}}\cos\left(\phi_{\beta}\right) & \mathscr{C}\\
 \mathscr{C} & \frac{2\tanh{r^{'}}}{1 - \tanh{r^{'}}^{2}}\cos\left(\phi_{\beta}\right) \label{ceq}
    \end{pmatrix}
\end{align}
%
\begin{equation}
\mathscr{C} =   2(\cosh{r'})^{-2} \sum_{N} (N+1) (\tanh{r'} )^{2N+1}\sin\left(\phi_\beta\right) \label{curly c}
\end{equation}
and $C = C^{T}$. 
The analytical expression for the unitary averaging model in the low noise limit assuming the individual random phases $\phi_{j}$ are independent Gaussian parameters with mean zero and variance $v$ can be approximated as
\begin{align}
    \left\langle\tanh{r'}\right\rangle \approx& \Big(1 - \big(\frac{v}{2} - \frac{v}{2n}\big)\Big)\tanh{r} \\
    \left\langle\cos\left(\phi_{\beta}\right)\right\rangle \approx& \cos{\left(\sqrt{\frac{v}{n}}\right)}
\end{align}
where we have assumed $v\ll1$ and
\begin{align}
\left\langle\frac{2\tanh{r^{'}}}{1 - \tanh{r^{'}}^{2}}\cos\left(\phi_{\beta}\right)\right\rangle & \approx \frac{2\left\langle\tanh{r^{'}}\right\rangle}{1 - \left\langle\tanh{r^{'}}\right\rangle^{2}}\left\langle\cos\left(\phi_{\beta}\right)\right\rangle\\
\langle \mathscr{C} \rangle & \approx 0
\end{align}

In summary, we introduced a unitary averaging (UA) scheme with continuous variables that incorporates vacuum detection \cite{swain2024improving}. This approach effectively mitigates the impact of phase noise within the channel. Interestingly, our results demonstrated more than an order of magnitude reduction in noise, along with improvements in squeezing, purity, and entanglement levels in practical systems. Moving forward, we will extend the UA scheme to a multi-mode channel with arbitrary interferometer noise. In the following section, we begin by examining the two-mode channel.

\section{BRIEF REVIEW OF THE PHASE-SPACE FORMALISM}
\label{BRIEF REVIEW OF THE PHASE-SPACE FORMALISM}
As outlined in the main text, in order to analyse the continuous-variable quantum key distribution protocol, we employ the phase-space formalism \cite{BAC19,weedbrook2012gaussian, ferraro2005gaussian, serafini2023quantum, olivares2012quantum}. The following subsections are intended merely as a concise review of the relevant formalism to support the results presented in the main text.

\subsection{Mode and quadrature operators}
We consider bosonic systems composed of $N$ harmonic oscillators, each associated with annihilation and creation operators $\hat{a}_{j}$ and $\hat{a}^{\dag}_{j}$, where $j = 1, \dots, N$. The operators $\hat{a}_{j}$ will also be referred to as mode operators. Throughout, we set $\hbar = 2$ and adopt the canonical commutation relations.
\begin{align}
    [\hat{a}_{j}, \hat{a}_{k}] & = [\hat{a}^{\dag}_{j}, \hat{a}^{\dag}_{k}] = 0,\\
    [\hat{a}_{j}, \hat{a}^{\dag}_{k}] & = \delta_{jk}
\end{align}

We define Hermitian position- and momentum-like operators for each mode
\begin{align}
    \hat{x}_{j} & = (\hat{a}_{j} + \hat{a}^{\dag}_{j} ) \\
    \hat{p}_{j} & = i (  \hat{a}^{\dag}_{j} - \hat{a}_{j} )
\end{align}
with the canonical commutator
\begin{align}
    [\hat{x}_{j}, \hat{p}_{k}] & = 2i\delta_{jk}
\end{align}

Following the convention in quantum optics, we also refer to $\hat{x}_{j}$ and $\hat{p}_{j}$ as quadrature operators. For convenience, we collect the $2N$ quadrature operators corresponding to the $N$ modes into a single column vector, which we define as

\begin{align}
    \hat{r } = \begin{pmatrix}
        \hat{x}_{1} \\
        \hat{p}_{1} \\
        \vdots \\
        \hat{x}_{N} \\
        \hat{p}_{N}
    \end{pmatrix}
\end{align}
The canonical commutation relations can then be written
\begin{align}
    [\hat{r}_{j}, \hat{r}_{k}] & = \Omega_{jk}
\end{align}
where $\Omega$ is the symplectic matrix defined by

\begin{align}
    \Omega = \bigoplus_{j=1}^{2N} \begin{pmatrix}
        0 & 1 \\
        -1 & 0
    \end{pmatrix} = \mathbb{I}_{N} \otimes  \begin{pmatrix}
        0 & 1 \\
        -1 & 0
    \end{pmatrix}
\end{align}

where $\mathbb{I}_{N}$ is the $N \times N$ identity matrix.

\subsection{Covariance Matrices}
Given a quantum state $\rho$ of an $N$-mode system, the
corresponding covariance matrix $V$ has elements

\begin{align}
    V_{jk} = \langle \{ \hat{r}_{j}, \hat{r}_{k}\} \rangle -  \langle  \hat{r}_{j} \rangle \langle  \hat{r}_{k} \rangle
\end{align}

where $\{  ., . \}$ denotes the anticommutator and $\langle \hat{A}\rangle $ is the expectation value.
For a valid density matrix $\rho$, the covariance matrix $V$ is real, symmetric, and positive definite ($V > 0$), and it satisfies

\begin{align}
    V + i\Omega \ge 0
\end{align}

Note that the diagonal elements of $V$ represent the variances of the quadrature operators, whereas the non-zero off-diagonal elements correspond to the correlations between different quadratures.

\section{CALCULATION OF THE VARIOUS PROTOCOL}
\label{CALCULATION FROM VARIOUS PROTOCOL}

As discussed in the main text, we conduct the security analysis by leveraging the optimality of Gaussian attacks \cite{pirandola2017fundamental}. When Alice and Bob share a non-Gaussian state $\rho$, a lower bound on the exact SKR can be obtained by considering a corresponding Gaussian protocol in which they share a Gaussian state $\rho_{G}$ possessing the same CM as $\rho$. In this Appendix, we derive the CM for the protocols presented in the main text. To achieve this, we employ the input-output formalism along with the Kraus operator representation of quantum states.

\subsection{KRAUS OPERATOR REPRESENTATION}

In a lossy thermal-noise channel, the input state $\rho_{0}$ interacts with a thermal state $\rho_{th}$ through a beam splitter with transmissivity $\eta$. After the interaction, the environment is traced out, resulting in the channel’s output state. A possible Kraus operator representation can be obtained by decomposing the lossy thermal-noise channel into a pure-loss channel with transmissivity $\tau = \eta / G$, followed by a quantum-limited amplifier with gain $G = 1 + (1 - \eta)\bar{n} = \frac{\eta \epsilon}{2} + 1$, where $\eta$ represents the pure loss and $\epsilon$ denotes the excess noise \cite{ivan2011operator}. This decomposition yields the following Kraus representation for the lossy thermal-noise channel:

\begin{align}
   \mathbb{W}(\rho_{0}) = \sum^{\infty}_{k,l} \hat{B}_{k} \hat{A}_{l} \rho_{0} \hat{A}^{\dag}_{l}\hat{B}^{\dag}_{k} \label{thloss}
\end{align}
where $\hat{A}_{l}$ are the Kraus operators of the pure loss channel, while $\hat{B}_{k}$ are the Kraus operators of the quantum limited amplifier \cite{ivan2011operator} given by

\begin{align}
    \hat{A}_{l} & = \sqrt{\frac{(1- \tau)^{l}}{l!}} \tau^{\hat{n}/2} \hat{a}^{l} \\
    \hat{B}_{k} & = \sqrt{\frac{1}{k!}\frac{1}{G}(\frac{G-1}{G})^{k}}\hat{a}^{k} G^{-\hat{n}/2}
\end{align}
where $\hat{n}$ is the number operator. $\hat{a}$ and $\hat{a}^{\dag}$ are annihilation and creation operator respectively.

\subsection{CALCULATION OF THE CM WITH PHASE NOISE}
In this subsection, we detailed the calculation of a scenario detailed in section \ref{UA-Assisted CV QKD} and in fig.\ref{fig:CirPh}.
The $\rho_{0}$ of input TMSV is,
\begin{align}
   \ket{ \psi_{0} }& = (\cosh{r})^{-1}\sum_{n = 0}^{\infty} (-\tanh{r})^{n}\ket{n, n}\\
    \rho_{0} & = (\cosh{r})^{-2}\sum_{n = 0}^{\infty} \sum_{m = 0}^{\infty} (\tanh{r})^{n+m}\ket{n, n}\bra{m,m} \label{rhotmsv}
\end{align}

The phase shift noise defined as
\begin{align}
    R(\phi) & = \exp(i \phi \hat{n}), \: \hat{n} = \hat{a}^{\dag}\hat{a}
\end{align}
The noise we consider is fluctuations in phase shifter (PS) parameters, which are modelled as independent Gaussian noise on each mode drawn from the same distribution.
With the application of $R_{2}(\phi)$ on second mode of the EQ.\ref{rhotmsv},
\begin{align}
    \rho_{\text{PS}} =&  R_{2}(\phi)(\cosh{r})^{-2}\sum_{n = 0}^{\infty} \sum_{m = 0}^{\infty} (\tanh{r})^{n+m} \nonumber \\
    &\ket{n, n}\bra{m,m}R_{2}(\phi)^{\dag} \\
    =& (\cosh{r})^{-2}\sum_{n = 0}^{\infty} \sum_{n = 0}^{\infty}(\tanh{r})^{n+m} \exp(i \phi \hat{a}_{2}^{\dag} \hat{a}_{2}) \nonumber\\
    & \ket{n, n}\bra{m,m}\exp(-i \phi \hat{a}_{2}\hat{a}_{2}^{\dag}) \label{rhops}
\end{align}

With the thermal loss channel, using Eq.\ref{thloss} in Eq.\ref{rhops}
\begin{align}
    \mathbb{W}(\rho_{\text{PS}}) = \sum^{\infty}_{k,l} \hat{B}_{k} \hat{A}_{l} \rho_{\text{PS}} \hat{A}^{\dag}_{l}\hat{B}^{\dag}_{k} \label{thlossps}
\end{align}

Finally, we compute the covariance matrix (CM) associated with the state eq.\ref{thlossps}.
\begin{align}
    \mathbb{A}_{\text{PS}} = &  \Tr(\rho_{\text{PS}}q_{A}^{2}) \\
    \mathbb{B}_{\text{PS}} = &  \Tr(\rho_{\text{PS}}q_{B}^{2}) \\
    \mathbb{C}_{\text{PS}} = &  \Tr(\rho_{\text{PS}}q_{A}q_{B})
\end{align}

Accordingly, the CM is written
\begin{align}
   \Sigma^{(\text{PS})}_{AB} = \begin{pmatrix}
        \mathbb{A}_{\text{PS}} \mathbf{I}_{2} & \mathbb{C}_{\text{PS}} \mathbf{\sigma}_{2} \\
        \mathbb{C}_{\text{PS}} \mathbf{\sigma}_{2} & \mathbb{B}_{\text{PS}}\mathbf{I}_{2} \label{COVps}
    \end{pmatrix}
\end{align}

The CM (\ref{COVps}) in terms of $\eta$ and $\epsilon$,

\begin{align}
    \Sigma^{(\text{PS})}_{AB} \equiv \begin{pmatrix}
        \cosh{2r} \mathbf{I}_{2} & \sqrt{\eta} \sinh{2r} \cos{\phi}\mathbf{\sigma}_{2} \\
        \sqrt{\eta} \sinh{2r} \cos{\phi} \mathbf{\sigma}_{2} & \eta(\cosh{2r} + \frac{1- \eta}{\eta} + \epsilon)\mathbf{I}_{2} \label{COVpsfinal}
    \end{pmatrix}
\end{align}
where $r$ is the squeezing factor and $\eta$ is the pure loss.

\subsection{CALCULATION OF CM WITH UNITARY AVERAGING}
\label{AliceUA}
In this subsection, we detail the calculation of results explained in section \ref{UA-Assisted CV QKD} and in fig.\ref{fig:CIRUA}.

The input state reads,
\begin{align}
    \psi_{1} = & \ket{ \psi_{0} }\ket{0} \\
    = &  (\cosh{r})^{-1}\sum_{n = 0}^{\infty} (-\tanh{r})^{n}\ket{n, n}_{a1,b1}\ket{0}_{a2} \\
    \rho_{1} = & (\cosh{r})^{-2}\sum_{n = 0}^{\infty} \sum_{m= 0}^{\infty} (\tanh{r})^{n+m} \nonumber \\
    &\ket{n, n, 0}_{a1,b1,a2}\bra{m,m, 0}_{a1,b1,a2} \\
    = &  (\cosh{r})^{-2}\sum_{n = 0}^{\infty} \sum_{m= 0}^{\infty} (\tanh{r})^{n+m} (\hat{a}^{\dag}_{1})^{n}(\hat{b}^{\dag}_{1})^{n} \nonumber \\
    & \ket{0,0,0}_{a1,b1,a2}\bra{0,0,0}_{a1,b1,a2}(\hat{a}^{\dag}_{1})^{m}(\hat{b}^{\dag}_{1})^{m} \label{rhoua1}
\end{align}

The unitary transformation with loss and excess noise 
\begin{align}
     \hat{\mathbf{U}} = & \hat{B}^{(50:50)}_{b1a2}\hat{R(\phi_{b1})}\mathbb{W}_{b1}\hat{R(\phi_{a2})}\mathbb{W}_{a2}\hat{B}^{(50:50)\dag}_{b1a2} 
\end{align}
where $\hat{B}^{(50:50)}_{b1a2}$ is the $50:50$ beamsplitter between modes $b1$ and $a2$. $\mathbb{W}_{b1}$ and $\mathbb{W}_{a2}$ are thermal loss channel on modes  $b1$ and $a2$ respectively. Similarly, $\hat{R(\phi_{b1})}$ and $\hat{R(\phi_{a2})}$ are phase shifters on mode $b1$ and $a2$ respectively. As $\mathbb{W}_{b1}$ and $\mathbb{W}_{a2}$ commute with beamsplitters and phase shifters, we can consider applying the the thermal loss channel after the decoding step.

Now, we define UA transformation as
\begin{align}
    \hat{\mathbf{U}}_{\text{UA}} = & \hat{B}^{(50:50)}_{b1a2}\hat{R(\phi_{b1})}\hat{R(\phi_{a2})}\hat{B}^{(50:50)\dag}_{b1a2} \\
    = & \exp(\frac{\pi}{4}(\hat{b1}^{\dag}\hat{a2} - \hat{b1}\hat{a2}^{\dag} ))\exp(i \phi_{b1}  \hat{b1}^{\dag} \hat{b1} ) \nonumber \\
    & \exp(i \phi_{a2}  \hat{a2}^{\dag} \hat{a2} ) \exp(\frac{\pi}{4}(\hat{b1}\hat{a2}^{\dag} - \hat{b1}^{\dag}\hat{a2} )) \label{UABSPS}
\end{align}

In the Heisenberg picture, the annihilation operators are transformed via the linear unitary Bogoliubov transformation as
\begin{align}
    \begin{pmatrix}
       \hat{b1}'\\
      \hat{a2}'
    \end{pmatrix} \to & \frac{1}{2}\begin{pmatrix}
        1 & 1 \\
        -1 & 1
    \end{pmatrix}\begin{pmatrix}
      e^{(i \phi_{b1})} & 0 \\
        0 & e^{(i \phi_{a2})}
    \end{pmatrix} \begin{pmatrix}
        1 & -1 \\
        1 & 1
    \end{pmatrix}\begin{pmatrix}
       \hat{b1}\\
      \hat{a2}
    \end{pmatrix}\\
    \to & \frac{1}{2}\begin{pmatrix}
        e^{(i \phi_{b1})} +  e^{(i \phi_{a2})} &  e^{(i \phi_{a2})} - e^{(i \phi_{b1})} \\
        e^{(i \phi_{a2})} - e^{(i \phi_{b1})} & e^{(i \phi_{b1})} +  e^{(i \phi_{a2})}
    \end{pmatrix}\begin{pmatrix}
       \hat{b1}\\
      \hat{a2}
    \end{pmatrix} \\
   \hat{b1}'  \to & \frac{1}{2}\big( ( e^{(i \phi_{b1})} +  e^{(i \phi_{a2})})\hat{b1} + (e^{(i \phi_{a2})} - e^{(i \phi_{b1})} )\hat{a2}\big) \label{b1ua}\\
  \hat{a2}' \to & \frac{1}{2}\big((e^{(i \phi_{a2})} - e^{(i \phi_{b1})} ) \hat{b1} + ( e^{(i \phi_{b1})} +  e^{(i \phi_{a2})})\hat{a2}\big)   
\end{align}

Applying $\hat{\mathbf{U}}_{\text{UA}}$ to the Eq.\ref{rhoua1}
\begin{align}
    \rho_{\text{UA}} \to & \hat{\mathbf{U}}_{\text{UA}}\rho_{1}\hat{\mathbf{U}}^{\dag}_{\text{UA}} \\
\rho_{\text{UA}} = &    (\cosh{r})^{-2}\sum_{n = 0}^{\infty} \sum_{m= 0}^{\infty} (\tanh{r})^{n+m} \hat{\mathbf{U}}_{\text{UA}}(\hat{a}^{\dag}_{1})^{n}(\hat{b}^{\dag}_{1})^{n} \nonumber \\
    & \ket{0,0,0}_{a1,b1,a2}\bra{0,0,0}_{a1,b1,a2}(\hat{a}^{\dag}_{1})^{m}(\hat{b}^{\dag}_{1})^{m}\hat{\mathbf{U}}^{\dag}_{\text{UA}}\\
    = &    (\cosh{r})^{-2}\sum_{n = 0}^{\infty} \sum_{m= 0}^{\infty} (\tanh{r})^{n+m} (\hat{a}^{\dag}_{1})^{n}(\hat{b}^{'\dag}_{1})^{n} \nonumber \\
    & \ket{0,0,0}_{a1,b1,a2}\bra{0,0,0}_{a1,b1,a2}(\hat{a}^{\dag}_{1})^{m}(\hat{b}^{'\dag}_{1})^{m}\hat{\mathbf{U}}^{\dag}_{\text{UA}}\\
    = &  (\cosh{r})^{-2}\sum_{n = 0}^{\infty} \sum_{m= 0}^{\infty} (\tanh{r})^{n+m} (\hat{a}^{\dag}_{1})^{n} \nonumber \\
     & \Big(\frac{1}{2}\big( ( e^{(i \phi_{b1})} +  e^{(i \phi_{a2})})\hat{b1} + (e^{(i \phi_{a2})} - e^{(i \phi_{b1})} )\hat{a2}\big)\Big)^{n} \nonumber \\
    & \ket{0,0,0}_{a1,b1,a2}\bra{0,0,0}_{a1,b1,a2}(\hat{a}^{\dag}_{1})^{m} \nonumber \\
    &\Big(\frac{1}{2}\big( ( e^{(i \phi_{b1})} +  e^{(i \phi_{a2})})\hat{b1} + (e^{(i \phi_{a2})} - e^{(i \phi_{b1})} )\hat{a2}\big)\Big)^{m} \label{rhoua2}
\end{align}

With the thermal loss channel, exploiting Eq.\ref{thloss} in Eq.\ref{rhoua2}
\begin{align}
    \mathbb{W}(\rho_{\text{UA}}) = & \sum^{\infty}_{kb1,lb1} \sum^{\infty}_{ka2,la2} \hat{B}_{kb1} \hat{A}_{lb1}  \hat{B}_{ka2} \hat{A}_{la2}  \nonumber \\
    &\rho_{\text{UA}} \hat{A}^{\dag}_{la2}\hat{B}^{\dag}_{ka2} \hat{A}^{\dag}_{lb1}\hat{B}^{\dag}_{kb1}  \\
    = &  (\cosh{r})^{-2}\sum_{n = 0}^{\infty} \sum_{m= 0}^{\infty} \sum^{\infty}_{kb1,lb1} \sum^{\infty}_{ka2,la2} \nonumber\\
   & \hat{B}_{kb1} \hat{A}_{lb1} \hat{B}_{ka2} \hat{A}_{la2}  (\tanh{r})^{n+m} (\hat{a}^{\dag}_{1})^{n} \nonumber \\
     & \Big(\frac{1}{2}\big( ( e^{(i \phi_{b1})} +  e^{(i \phi_{a2})})\hat{b1} + (e^{(i \phi_{a2})} - e^{(i \phi_{b1})} )\hat{a2}\big)\Big)^{n} \nonumber \\
    & \ket{0,0,0}_{a1,b1,a2}\bra{0,0,0}_{a1,b1,a2}(\hat{a}^{\dag}_{1})^{m} \nonumber \\
    &\Big(\frac{1}{2}\big( ( e^{(i \phi_{b1})} +  e^{(i \phi_{a2})})\hat{b1} + (e^{(i \phi_{a2})} - e^{(i \phi_{b1})} )\hat{a2}\big)\Big)^{m} \nonumber\\
   & \hat{A}^{\dag}_{la2}\hat{B}^{\dag}_{ka2} \hat{A}^{\dag}_{lb1}\hat{B}^{\dag}_{kb1} \label{thlossUA}
\end{align}
After applying the vacuum projection operator on mode $a3$, the normalised output state  is
\begin{align}
    \rho^{'}_{\text{UA}} = & \frac{(\mathbf{I}_{a1} \otimes \mathbf{I}_{b1} \otimes \bra{0})\mathbb{W}(\rho_{\text{UA}}) (\mathbf{I}_{a1} \otimes \mathbf{I}_{b1} \otimes \ket{0})} {\Tr\Big((\mathbf{I}_{a1} \otimes \mathbf{I}_{b1} \otimes \bra{0})\mathbb{W}(\rho_{\text{UA}})\Big)} \label{uarhonormalised}
\end{align}
Normalisation constant represents the success probability of the vacuum detection.

Finally, we compute the CM associated with the state (\ref{uarhonormalised}).
\begin{align}
    \mathbb{A}_{\text{UA}} = &  \Tr(\rho^{'}_{\text{UA}}q_{a1}^{2}) \\
    \mathbb{B}_{\text{UA}} = &  \Tr(\rho^{'}_{\text{UA}}q_{b1}^{2}) \\
    \mathbb{C}_{\text{UA}} = &  \Tr(\rho^{'}_{\text{UA}}q_{a1}q_{b1})
\end{align}

Accordingly, the CM is written
\begin{align}
   \Sigma^{(\text{UA})}_{a1b1} = \begin{pmatrix}
        \mathbb{A}_{\text{UA}} \mathbf{I}_{2} & \mathbb{C}_{\text{UA}} \mathbf{\sigma}_{2} \\
        \mathbb{C}_{\text{UA}} \mathbf{\sigma}_{2} & \mathbb{B}_{\text{UA}}\mathbf{I}_{2} \label{COVUA}
    \end{pmatrix}
\end{align}
where
\begin{align}
\mathbf{Z} =  & \big(\frac{e^{i \phi_{b1}} + e^{i \phi_{a2}} }{2}\big); \: \phi_{j} \sim \mathcal{N}(0, \sigma^2) \\
    \mathbb{A}_{\text{UA}} \equiv & \cosh{2r}\sqrt{\Big(\Re(\mathbf{Z})\Big)^{2} + \Big(\Im(\mathbf{Z})\Big)^{2}} \\
    \mathbb{B}_{\text{UA}} \equiv & \eta \Big(\mathbb{A}_{\text{UA}} + \frac{1- \eta}{\eta} + \epsilon\Big) \\
    \mathbb{C}_{\text{UA}} \equiv & \sqrt{\eta}\sinh{2r}\sqrt{\Big(\Re(\mathbf{Z})\Big)^{2} + \Big(\Im(\mathbf{Z})\Big)^{2}} \nonumber \\ 
   & \cos{\Big(\arctan{\frac{\Im(\mathbf{Z})}{\Re(\mathbf{Z})}}\Big)}
\end{align}
where $\Re(\mathbf{Z})$ and $\Im(\mathbf{Z})$ is the real and imaginary part of $\mathbf{Z}$.

\subsection{EMPLOYING QUANTUM SCISSOR AND UNITARY AVERAGING}
\label{EMPLOYING QUANTUM SCISSOR AND
UNITARY AVERAGING}

Alice prepares three modes ($a1,\: a2,\: a3$). She applies an encoding beamsplitter between modes ($a2$) and ($a3$) and then sends the encoded modes to Charlie’s station. Likewise, Bob prepares three modes ($b1,\: b2, \: b3$), applies an encoding beamsplitter between ($b1$) and ($b3$), and forwards the encoded modes to Charlie. Charlie performs vacuum post-selection on modes ($a3$) and ($b3$), then interferes modes ($a2$) and ($b1$) and measures the result with photon-number-resolving detectors.

From subsection \ref{AliceUA}, following the vacuum detection on mode ($a3$), Alice’s unitary-averaged state is given by

\begin{align}
    \rho^{'}_{\text{A}} = & \frac{(\mathbf{I}_{a1} \otimes \mathbf{I}_{a2} \otimes \bra{0})\mathbb{W}(\rho_{\text{A}}) (\mathbf{I}_{a1} \otimes \mathbf{I}_{a2} \otimes \ket{0})} {\Tr\Big((\mathbf{I}_{a1} \otimes \mathbf{I}_{a2} \otimes \bra{0})\mathbb{W}(\rho_{\text{UA}})\Big)} \label{uarhonormalised}
\end{align}

Bob prepares three modes in the Fock states $\ket{1}_{b1}$, $\ket{0}_{b2}$, and $\ket{1}_{b3}$ respectively.  He mixes them at a beam splitter with
transmissivity $\mathbb{T} = \frac{1}{1 + g^{2}}$. It fixes the gain associated with the NLA; that for low-amplitude coherent signals reads $ g = \sqrt{\frac{1 - \mathbb{T}}{\mathbb{T}}}$

\begin{align}
    \ket{\psi}_{\text{B}} =& \ket{1}_{b1}\ket{0}_{b2}\ket{0}_{b3} \\
    \rho_{\text{B}} = &  \ket{1}_{b1}\ket{0}_{b2}\ket{0}_{b3} \bra{1}_{b1}\bra{0}_{b2}\bra{0}_{b3} \label{BOBINRHO}
\end{align}

The beamsplitter with transmissivity $\mathbb{T} = \frac{1}{1 + g^{2}}$ changes the modes $b1$ and $b2$ as

\begin{align}
    \begin{pmatrix}
       \hat{b1}'\\
      \hat{b2}'
    \end{pmatrix} \to & \begin{pmatrix}
        \sqrt{\mathbb{T}} & \sqrt{1 -\mathbb{ T}}\\
        -\sqrt{1 -\mathbb{ T}} & \sqrt{\mathbb{T}}
    \end{pmatrix} \begin{pmatrix}
       \hat{b1}\\
      \hat{b2}
    \end{pmatrix} \label{NLABS}
\end{align}

Applying Eq.\ref{NLABS} to the state (\ref{BOBINRHO}),
\begin{align}
    \rho_{\text{B}}= &  \Big(\sqrt{\mathbb{T}}\hat{b1}^{\dag} + \sqrt{1 -\mathbb{ T}}\hat{b2}^{\dag}\Big)\ket{0}_{b1}\ket{0}_{b2}\ket{0}_{b3} \nonumber \\
   & \bra{0}_{b1}\bra{0}_{b2}\bra{0}_{b3}\Big(\sqrt{\mathbb{T}}\hat{b1}^{\dag} + \sqrt{1 -\mathbb{ T}}\hat{b2}^{\dag}\Big)  
\end{align}

After UA, using Eq.\ref{UABSPS},
\begin{align}
    \hat{b1}'  \to & \frac{1}{2}\big( ( e^{(i \phi_{b1})} +  e^{(i \phi_{b3})})\hat{b1} + (e^{(i \phi_{b3})} - e^{(i \phi_{b1})} )\hat{b3}\big) 
\end{align}

\begin{align}
    \rho_{\text{B}} = &  \Big(\sqrt{\mathbb{T}}\hat{b1}^{'\dag} + \sqrt{1 -\mathbb{ T}}\hat{b2}^{\dag}\Big)\ket{0}_{b1}\ket{0}_{b2}\ket{0}_{b3} \nonumber \\
   & \bra{0}_{b1}\bra{0}_{b2}\bra{0}_{b3}\Big(\sqrt{\mathbb{T}}\hat{b1}^{'\dag} + \sqrt{1 -\mathbb{ T}}\hat{b2}^{\dag}\Big)  \\
   = &  \Big(\sqrt{\mathbb{T}}\Big(\frac{1}{2}\big( ( e^{(i \phi_{b1})} +  e^{(i \phi_{b3})})\hat{b1} + (e^{(i \phi_{b3})} - e^{(i \phi_{b1})} )\hat{b3}\big)\Big) \nonumber\\
 &  + \sqrt{1 -\mathbb{ T}}\hat{b2}^{\dag}\Big)\ket{0}_{b1}\ket{0}_{b2}\ket{0}_{b3} \bra{0}_{b1}\bra{0}_{b2}\bra{0}_{b3}\nonumber \\   
 & \Big(\sqrt{\mathbb{T}}\Big(\frac{1}{2}\big( ( e^{(i \phi_{b1})} +  e^{(i \phi_{b3})})\hat{b1} + (e^{(i \phi_{b3})} - e^{(i \phi_{b1})} )\hat{b3}\big)\Big) \nonumber \\
 & + \sqrt{1 -\mathbb{ T}}\hat{b2}^{\dag}\Big)
\end{align}

With the thermal loss channel (\ref{thloss}),
\begin{align}
    \mathbb{W}(\rho_{\text{B}}) = & \sum^{\infty}_{kb1,lb1} \sum^{\infty}_{kb3,lb3} \hat{B}_{kb1} \hat{A}_{lb1}  \hat{B}_{kb3} \hat{A}_{lb3}  \nonumber \\
    &\rho_{\text{B}} \hat{A}^{\dag}_{lb3}\hat{B}^{\dag}_{kb3} \hat{A}^{\dag}_{lb1}\hat{B}^{\dag}_{kb1}  \\
    = & \sum^{\infty}_{kb1,lb1} \sum^{\infty}_{kb3,lb3}\hat{B}_{kb1} \hat{A}_{lb1}  \hat{B}_{kb3} \hat{A}_{lb3} \nonumber\\
    &\Big(\sqrt{\mathbb{T}}\Big(\frac{1}{2}\big( ( e^{(i \phi_{b1})} +  e^{(i \phi_{b3})})\hat{b1} + (e^{(i \phi_{b3})} - e^{(i \phi_{b1})} )\hat{b3}\big)\Big) \nonumber\\
    &  + \sqrt{1 -\mathbb{ T}}\hat{b2}^{\dag}\Big)\ket{0}_{b1}\ket{0}_{b2}\ket{0}_{b3} \bra{0}_{b1}\bra{0}_{b2}\bra{0}_{b3}\nonumber \\   
 & \Big(\sqrt{\mathbb{T}}\Big(\frac{1}{2}\big( ( e^{(i \phi_{b1})} +  e^{(i \phi_{b3})})\hat{b1} + (e^{(i \phi_{b3})} - e^{(i \phi_{b1})} )\hat{b3}\big)\Big) \nonumber \\
 & + \sqrt{1 -\mathbb{ T}}\hat{b2}^{\dag}\Big)\hat{A}^{\dag}_{lb3}\hat{B}^{\dag}_{kb3} \hat{A}^{\dag}_{lb1}\hat{B}^{\dag}_{kb1}
\end{align}

After applying the vacuum projection operator on mode $b3$, the normalised output state  is
\begin{align}
    \rho^{'}_{\text{B}} = & \frac{(\mathbf{I}_{b1} \otimes \mathbf{I}_{b2} \otimes \bra{0})\mathbb{W}(\rho_{\text{B}}) (\mathbf{I}_{b1} \otimes \mathbf{I}_{b2} \otimes \ket{0})} {\Tr\Big((\mathbf{I}_{b1} \otimes \mathbf{I}_{b2} \otimes \bra{0})\mathbb{W}(\rho_{\text{B}})\Big)} \label{Bobfinal}
\end{align}

Normalisation constant represents the success probability of the vacuum detection of Bob's mode at Charlie's station.

Now Charlie connected modes $a2$ and $b1$ via a balanced beamsplitter and performs photon number resolving detection.
\begin{align}
\rho_{\text{AB}} = & \hat{B}^{(50:50)}_{a2b1} \Big(\rho_{\text{A}}(a1, a2) \otimes \rho_{\text{B}}(b1, b2)\Big) \hat{B}^{(50:50)\dag}_{a2b1} \\
    \rho^{'}_{\text{AB}} = & \frac{(\mathbf{I}_{a1} \otimes \bra{0} \otimes \bra{1}\otimes \mathbf{I}_{b2} )\rho_{\text{AB}} (\mathbf{I}_{a1} \otimes \ket{0} \otimes \ket{1}\otimes \mathbf{I}_{b2} )} {\Tr\Big((\mathbf{I}_{a1} \otimes \bra{0} \otimes \bra{1}\otimes \mathbf{I}_{b2} )\rho_{\text{AB}} \Big)}
\end{align}

 The global success probability of the QS-based NLA is $P_{\text{QS}} = 2  \Tr\big((\mathbf{I}_{a1} \otimes \bra{0} \otimes \bra{1}\otimes \mathbf{I}_{b2} )\rho_{\text{AB}} \big) $.

Finally, we compute the CM associated with the state (\ref{Bobfinal}).
\begin{align}
    \mathbb{A}_{\text{AB}} = &  \Tr(\rho^{'}_{\text{AB}}q_{a1}^{2}) \\
    \mathbb{B}_{\text{AB}} = &  \Tr(\rho^{'}_{\text{AB}}q_{b2}^{2}) \\
    \mathbb{C}_{\text{AB}} = &  \Tr(\rho^{'}_{\text{AB}}q_{a1}q_{b2})
\end{align}

Accordingly, the CM is written
\begin{align}
   \Sigma^{(\text{AB})}_{a1b2} = \begin{pmatrix}
        \mathbb{A}_{\text{AB}} \mathbf{I}_{2} & \mathbb{C}_{\text{AB}} \mathbf{\sigma}_{2} \\
        \mathbb{C}_{\text{AB}} \mathbf{\sigma}_{2} & \mathbb{B}_{\text{AB}}\mathbf{I}_{2} 
    \end{pmatrix}
\end{align}

\bibliography{references}

\end{document}